\documentclass[trackchanges]{aastex701}
\usepackage[T1]{fontenc}
\usepackage[utf8]{inputenc}

\begin{document}
\begin{sloppypar}

\title{Enhanced Non-Thermal Line Broadening inside Coronal Cavities above Solar Prominences revealed by Spectral Imaging CoronaGraph}

\correspondingauthor{Hui Fu}
\email{fuhui@sdu.edu.cn}
\correspondingauthor{LiDong Xia}
\email{xld@sdu.edu.cn}

\author{Chenxi Huangfu}
\affiliation{Shandong Key Laboratory of Space Environment and Exploration Technology, Institute of Space Sciences, School of Space Science and Technology, Shandong University, Shandong, China}
\email{huangfucx1014@163.com}

\author[0000-0002-8827-9311]{Hui Fu}
\affiliation{Shandong Key Laboratory of Space Environment and Exploration Technology, Institute of Space Sciences, School of Space Science and Technology, Shandong University, Shandong, China}
\email{fuhui@sdu.edu.cn}

\author[0000-0003-4790-6718]{Bo Li}
\affiliation{Shandong Key Laboratory of Space Environment and Exploration Technology, Institute of Space Sciences, School of Space Science and Technology, Shandong University, Shandong, China}
\email{BBL@sdu.edu.cn}

\author[0000-0002-2358-5377]{ZhengHua Huang}
\affiliation{Institute of Science and Technology for Deep Space Exploration, Suzhou Campus, Nanjing University, Suzhou 215163, People's Republic of China}
\email{huangzh@nju.edu.cn}

\author[0000-0002-1631-9898]{MingZhe Sun}
\affiliation{Shandong Key Laboratory of Space Environment and Exploration Technology, Institute of Space Sciences, School of Space Science and Technology, Shandong University, Shandong, China}
\email{sunmingzhe@sdu.edu.cn}

\author{WeiXin Liu}
\affiliation{Shandong Key Laboratory of Space Environment and Exploration Technology, Institute of Space Sciences, School of Space Science and Technology, Shandong University, Shandong, China}
\email{liuwx@sdu.edu.cn}

\author{XiaoYu Yu}
\affiliation{Shandong Key Laboratory of Space Environment and Exploration Technology, Institute of Space Sciences, School of Space Science and Technology, Shandong University, Shandong, China}
\email{yxy@sdu.edu.cn}

\author[0000-0001-8938-1038]{LiDong Xia}
\affiliation{Shandong Key Laboratory of Space Environment and Exploration Technology, Institute of Space Sciences, School of Space Science and Technology, Shandong University, Shandong, China}
\email{xld@sdu.edu.cn}

\begin{abstract}

Coronal cavities, often associated with prominences, are crucial structures in understanding coronal heating and the eruption mechanism of Coronal Mass Ejections (CMEs). Previous studies have identified their lower density, higher temperature, and flux rope structures. However, spectroscopic observations are still relatively scarce. In this study, we utilize the newly developed Spectral Imaging Coronagraph (SICG), Chinese H$\alpha$ Solar Explorer (CHASE), and AIA/SDO to analyze the morphology, temperature, Doppler shift, and non-thermal velocity of two coronal cavities observed on November 13, 2024. We find that coronal cavities are distinctly visible in SICG \ion{Fe}{14} 5303~\AA\ and AIA 193~\AA, whereas they are nearly absent in SICG \ion{Fe}{10} 6374~\AA\ and AIA 171~\AA. The spectroscopic measurements show that the two coronal cavities display asymmetric, ring-like structures in the \ion{Fe}{14} 5303~\AA\ Doppler shift maps. The non-thermal velocities inside coronal cavities are significantly higher than those of the surrounding streamer areas. In addition, the core regions of coronal cavities, located directly above the prominences, exhibit the highest non-thermal velocities and Doppler velocities. Our results suggest the presence of waves and turbulence in coronal cavities, which are likely more intense than those in the adjacent streamer regions. We suggest that the interaction and exchange between the cold, dense prominence materials and the hot, low-density coronal materials are the main drivers of the waves and turbulence inside coronal cavities.

\end{abstract}


\section{Introduction}

Coronal cavities are crucial structures in understanding coronal heating and the eruption mechanism of CMEs. Coronal cavities generally surround quiescent prominences, and are long-lived structures that can remain stable for several days to even weeks before eruption \citep{Gibson2006}. Coronal cavities are dark, elliptical cross-sectional structures. Compared to surrounding streamers, the coronal cavity exhibits a lower density (approximately 30\%) and a higher temperature in the range of 1.5 to 2~MK \citep{Hudson1999, Fuller2008, Gibson2010, Habbal2010c, Gibson2015, BS2019}. Coronal cavities can be observed across various wavelengths. In soft X-rays, a bright and hot core has been observed below streamers \citep{Hudson1999, Hudson2000, Reeves2012}. Radio observations can detect coronal cavities on disk, characterized by radio depression associated with filaments \citep{Marque2002, Marque2004}. White-light eclipse observations show the presence of complex twisted helical structures and turbulent-looking features inside coronal cavities \citep{Habbal2010a, Habbal2010c, Habbal2014, Druckmuller2014, Habbal2021}. White-light observations are more effective for detecting large-scale cavities, while smaller cavities are more readily observed at EUV wavelengths \citep{Gibson2015}. The projection of coronal cavities on the plane of the sky in extreme ultraviolet images is typically well-described by an ellipse. A statistical analysis of 129 EUV coronal cavities shows that 93\% of them have a height greater than their width \citep{Forland2013}.

The coronal cavity is a region where cold, dense prominence materials interact with hot, low-density coronal materials, featuring fine structures and natural instabilities \citep{Habbal2010c, Habbal2014, Habbal2021, Druckmuller2014}. The frequent coexistence of coronal cavities and prominences suggests a connection between them \citep{Forland2013, BS2016}. \citet{Berger2011} observed the presence of dark and low-density bubbles in prominence. These bubbles contain plasma at 0.25--1.2~MK, which is 25--120 times hotter than the overlying prominence. Through Rayleigh-Taylor instability, the bubbles rise into the coronal cavity above the prominence. These observations show that coronal-temperature plasma can be injected into the coronal cavity from the prominence. \citet{Berger2012} observed a bright ``cloud'' in the central area of the coronal cavity. The height and temperature of the ``cloud'' decreased over time. As the coronal cavity darkens, the prominence gradually grows. The above observation suggests that prominence can be formed by condensation of hot plasma inside the coronal cavity. The horn is a coherent loop-like bright structure seen in the 171~\AA\ band, which is formed of cool plasma. This structure is a coronal feature related to prominence dynamics. It connects the prominence and the coronal cavity \citep{Schmit2013}. The simulation suggests that the horn is a sign of the development of hyperbolic flux tube (HFT) topology and tether-cutting reconnections in the current sheet formed along the HFT \citep{Fan2012}. \citet{Wang2016} observed that the rising parts of the prominence and the related horns gradually disappeared into the upper corona. One day later, a coronal cavity was observed. This shows that magnetic flux, mass, and magnetic energy were transferred from the prominence to the coronal cavity. The horn can act as a field-tracing structure. It shows the transfer of magnetic field and energy between the prominence and the coronal cavity.

Spectroscopic observations reveal structured velocity fields inside coronal cavities. A series of studies has utilized the Coronal Multi-channel Polarimeter (CoMP; \citealt{Tomczyk2008}) and the Extreme Ultraviolet Imaging Spectrometer (EIS; \citealt{Culhane2007}) to analyze coronal cavities above solar prominences. The Doppler shift characteristic inside coronal cavities was detected for the first time by \citet{Schmit2009}, using CoMP and EIS. They found both red-shift and blue-shift structures inside coronal cavities. The magnitude of Doppler velocities ranges from 5 to 10 km\,s$^{-1}$, with spatial scales of tens of megameters, and durations of at least one hour. The above results suggest that there exist stable material flows inside coronal cavities. \citet{BS2016} analyzed 66 days of observations of non-erupting coronal cavities observed by CoMP. They found that nested rings of Doppler shift are commonly observed within coronal cavities exhibiting alternating red-shift and blue-shift structures. The magnitude of Doppler velocities typically ranges from 4 to 8 km\,s$^{-1}$. The velocity range was found to correlate with the size of the coronal cavity, with wider coronal cavities exhibiting larger velocity ranges. In addition, the highest Doppler velocity is located at the center of coronal cavities, positioned just above the prominences. Generally, the magnetic field structure of the coronal cavity has been modeled as a flux rope \citep{Xia2014, Fan2018, Jenkins2021}. The nested ring structures of Doppler shift inside coronal cavities are regarded as strong evidence for the flux-rope model, as these structures represent plasma flows along helical magnetic field structures \citep{BS2013, Chen2018}.

Investigating spectral line broadening can enhance our understanding of fluctuations, turbulence, and energy dissipation processes inside coronal cavities. To the best of our knowledge, no studies have yet been conducted on spectral line broadening within coronal cavities. In this paper, we utilized spectroscopic observations from the Spectral Imaging Coronagraph (SICG) to study the spectroscopic properties of coronal cavities and surrounding streamers on November 13, 2024. Consistent with previous studies, SICG observations show that coronal cavities are distinctly visible in \ion{Fe}{14} 5303~\AA\ (temperature nearly 2~MK) images, whereas they are nearly absent in \ion{Fe}{10} 6374~\AA\ (temperature nearly 1~MK) images. The Doppler shift maps present ring-like structures inside coronal cavities. More importantly, we find that the non-thermal velocities inside coronal cavities are significantly higher than those of surrounding streamer areas. The core regions of coronal cavities, which lie directly above the prominences, exhibit the highest non-thermal velocities. We describe the instruments and data analysis in Section~\ref{sec:Instruments and Data Analysis}, present the results in Section~\ref{sec:Results}, and discuss the observational findings in Section~\ref{sec:Discussion}. In Section~\ref{sec:Summary}, we summarize the main results. Additional morphological details and temperatures of coronal cavities derived from DEM analysis are given in Appendix~\ref{appendix}.

\section{Instruments and Data Analysis}\label{sec:Instruments and Data Analysis}

In the present study, we analyze two coronal cavities that were observed simultaneously at the east limb at approximately 01:06~UT on November 13, 2024. The observational data come mainly from the Spectral Imaging Coronagraph (SICG), the H$\alpha$ Imaging Spectrograph (HIS; \citealt{2022SCPMA..6589605L, 2022SCPMA..6589603Q}) onboard the Chinese H$\alpha$ Solar Explorer (CHASE; \citealt{2019RAA....19..165L, 2022SCPMA..6589602L}), and the Atmospheric Imaging Assembly \citep[AIA;][]{Lemen2012} onboard the Solar Dynamics Observatory \citep[SDO;][]{Pesnell2012}. On this day, SICG acquired 216 data sets in each of the 5303~\AA\ and 6374~\AA\ wavelength channels, covering the time period from 01:01 to 09:51~UT, with a temporal resolution of 2 minutes. The spatial resolution is $2\arcsec$ pix$^{-1}$, with a field of view (FOV) of $1.05$--$2\,R_{\sun}$.

SICG is designed for capturing images of the E-corona emission spectral line within the solar atmosphere. The instrument observes in two spectral lines: \ion{Fe}{10} 6374~\AA\ and \ion{Fe}{14} 5303~\AA. It is designed to conduct quasi-simultaneous observations at five spectral locations: 6372.00~\AA, 6373.70~\AA, 6374.35~\AA, 6375.00~\AA, 6377.00~\AA\ for \ion{Fe}{10} 6374~\AA\ and 5301.00~\AA, 5302.00~\AA, 5302.65~\AA, 5303.30~\AA, 5306.00~\AA\ for \ion{Fe}{14} 5303~\AA. In general, SICG takes approximately 40 seconds to acquire a set of data covering five spectral points. The filter wheel takes about 5 seconds to move to the next spectral location, and the exposure time is several seconds at each spectral location. For quasi-static coronal structures, an observation duration of 40 seconds can be regarded as a quasi-simultaneous observation.

The initial steps for data processing include removal of camera background noise, flat-field correction, and photometric calibration. The images at 6372.00~\AA\ and 6377.00~\AA\ (5301.00~\AA\ and 5306.00~\AA) are used as references for \ion{Fe}{10} 6374~\AA\ (\ion{Fe}{14} 5303~\AA) to subtract the sky background, K-corona background, and some stray light from the instrument. Then, an analytical solution is calculated using three middle points to derive the line profile parameters. Finally, the E-corona intensity, Doppler velocity, and line width can be derived from the line profile parameters (see \citet{Tian2013} for details). The non-thermal velocity was then calculated from the line width under the assumption that the ion temperature equals the electron temperature (see \citet{Xia2003} for details). We apply the Normalized Radial Graded Filter (NRGF; \citealt{Morgan2006}) to process the central-wavelength images of the two SICG bands at 5302.65~\AA\ and 6374.35~\AA. This method removes the brightness variation with height in coronal structures, allowing a more complete display of the coronal cavity structure.

The H$\alpha$ (6559.7--6565.9~\AA) images were taken by CHASE/HIS, which was running in raster scanning mode. The H$\alpha$ observations have been corrected, including dark-field correction, flat-field correction, slit image curvature correction, wavelength and intensity calibration, and coordinate transformation \citep{2022SCPMA..6589603Q}. The Level 1 CHASE data are available through the website of the Solar Science Data Center of Nanjing University\footnote{\url{https://ssdc.nju.edu.cn}}. The spatial resolution is $1.04\arcsec$ pix$^{-1}$ and the temporal cadence is about 60~s.

For AIA, we utilized image data from the 171 \AA\ and 193 \AA\ wavelength channels, with a spatial resolution of $0.6\arcsec$ pix$^{-1}$ and a temporal cadence of 12~s. All AIA images are processed using the \texttt{aia\_rfilter.pro} procedure in SSW to enhance contrast, providing a clearer view of the coronal cavity structures. The temperatures of coronal cavities are derived using the Differential Emission Measure (DEM) method developed by \citet{Cheung2015} and \citet{Su2018}. The DEM-weighted average temperature of coronal cavities is adopted in the present study.

\section{Results} \label{sec:Results}

\begin{figure*}[ht!]
\plotone{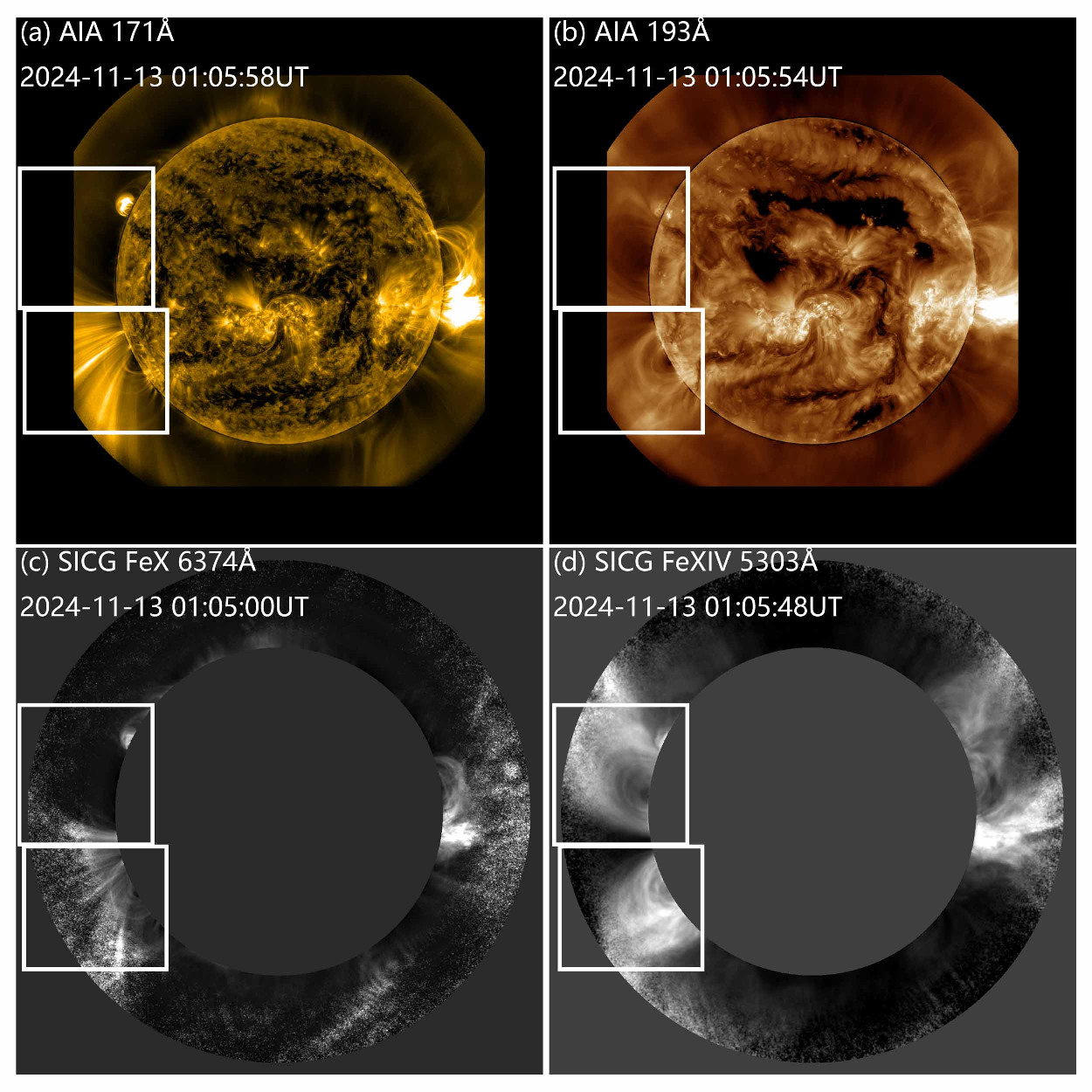}
\caption{Coronal images taken by SICG and SDO/AIA on November 13, 2024. The AIA 171 \AA\ and AIA 193 \AA\ images are shown in panels (a) and (b). The \ion{Fe}{10} 6374~\AA\ and  \ion{Fe}{14} 5303~\AA\ central wavelength images are shown in panels (c) and (d). The white boxes highlight the position of the two coronal cavities. The images have been processed to enhance contrast and highlight fine structures inside the coronal cavity and surrounding regions.}
\label{fig:1}
\end{figure*}

Figure~\ref{fig:1} shows coronal images taken by SICG and full-disk images taken by SDO/AIA at approximately 01:06~UT on November 13, 2024. The top panels present full-disk images from AIA, while the bottom panels show off-limb coronal images from SICG. The comparison demonstrates that both instruments observed highly consistent coronal structures at similar temperatures. The left panels display the AIA 171~\AA\ image (Figure~\ref{fig:1} (a)) and the SICG \ion{Fe}{10} 6374~\AA\ image (Figure~\ref{fig:1} (c)). These two images exhibit good agreement in coronal structures. For example, both images show a bright active region in the low-latitude area on the western limb and a fan-shaped structure extending outward in the mid- to low-latitude region on the eastern limb. The right panels show the AIA 193~\AA\ image (Figure~\ref{fig:1} (b)) and the SICG \ion{Fe}{14} 5303~\AA\ image (Figure~\ref{fig:1} (d)). These images also exhibit good structural consistency. A bright active region is again observed in the low-latitude western area. More importantly, both instruments reveal two coronal cavities in the mid- to low-latitude regions on the eastern limb, as marked by the white rectangles. The coronal cavities are distinctly visible in AIA 193 Å (Figure~\ref{fig:1} (b)) and SICG Fe XIV 5303 Å images (Figure~\ref{fig:1} (d)). The morphologies of coronal cavities, including their size, position, and surrounding structures, appear consistent between the AIA and SICG observations. The consistency confirms the reliability of the SICG data.

\begin{figure*}[ht!]
\plotone{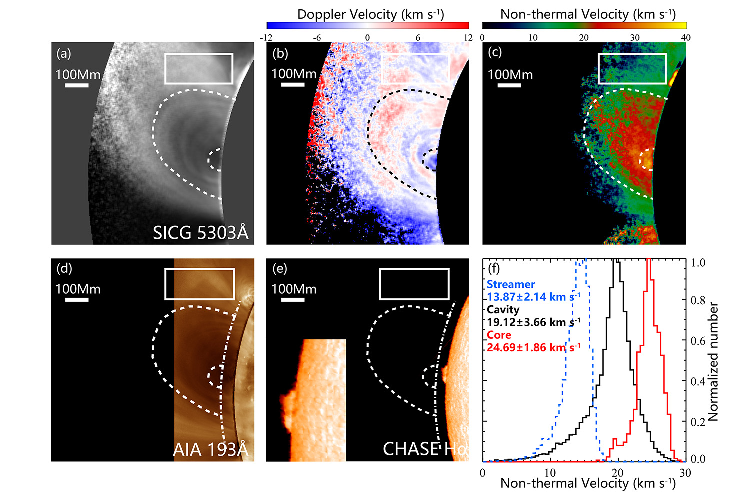}
\caption{The spectral properties of the north-east coronal cavity. The central wavelength intensity, Doppler velocity, and non-thermal velocity maps of the north-east coronal cavity derived from SICG \ion{Fe}{14} 5303~\AA\ are shown in panels (a), (b), and (c). The AIA 193 \AA \ image is shown in panel (d). The HIS H$\alpha$ line-center image is shown in panel (e), illustrating the spatial relationship between the coronal cavity, core region, and prominence. In panels (d) and (e), the positions of the SICG occulter are indicated by white dash-dotted lines. In panels (a), (b), (c), (d), and (e), the boundaries of the coronal cavities and core regions are marked with dashed lines, and the white boxes denote streamer regions near the coronal cavities. The histograms of non-thermal velocity distributions for streamers (blue), whole cavities (black, including the core), and core regions (red) are shown in panels (f).}

\label{fig:cavity1}
\end{figure*}

\begin{figure*}[ht!]
\plotone{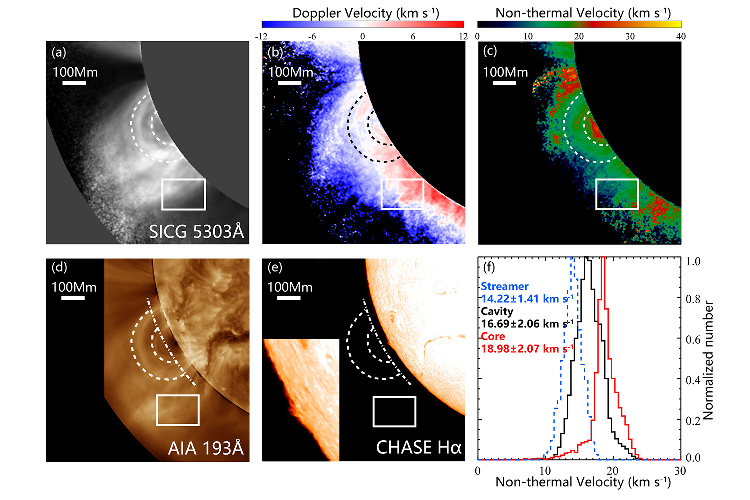}
\caption{Same as Figure 2 but for the south-east coronal cavity.}
\label{fig:cavity2}
\end{figure*}


We examine two coronal cavities in more detail in Figure~\ref{fig:cavity1} and Figure~\ref{fig:cavity2}. The boundaries of coronal cavities and core regions are
marked with dashed lines in panels (a), (b), (c), (d) and (e). The darkest semicircular regions are present at their centers, which we define as the core regions. The north-east coronal cavity appears as a semi-elliptical structure in the \ion{Fe}{14} 5303~\AA\ image (Figure~\ref{fig:cavity1} (a)) and AIA 193~\AA\ image (Figure~\ref{fig:cavity1} (d)). The south-east coronal cavity (Figures~\ref{fig:cavity2} (a) and (d)) shows a similar semicircular structure, but is smaller. The additional morphological details and temperatures of coronal cavities derived from DEM analysis are provided in Appendix~\ref{appendix}.

The Doppler velocity maps of the north-east and south-east coronal cavities derived from SICG \ion{Fe}{14} 5303~\AA\ are shown in Figure~\ref{fig:cavity1} (b) and Figure~\ref{fig:cavity2} (b). It is clear that ring-like structures appear inside the coronal cavities. For the north-east coronal cavity, the Doppler velocities inside coronal cavity range from $-12$ to $12~\mathrm{km~s^{-1}}$. The center of the ring-like pattern appears as a semicircular structure. This structure roughly coincides with the marked cavity core region and lies directly above the prominence. It shows the maximum Doppler shift of $-11~\mathrm{km~s^{-1}}$. Surrounding this core, alternating layers of blue- and red-shifted semicircular patterns are observed. These patterns are not perfectly symmetric. For the south-east coronal cavity, the maximum Doppler shift of $10~\mathrm{km~s^{-1}}$ is found inside the core region and lies directly above the prominence. Similar layered blue- and red-shifted structures are present. However, they are less distinct than those in the north-east coronal cavity.

The non-thermal velocity maps of the north-east and south-east coronal cavities derived from SICG \ion{Fe}{14} 5303~\AA\ are shown in Figure~\ref{fig:cavity1} (c) and Figure~\ref{fig:cavity2} (c). The histograms of non-thermal velocity distributions for streamers (denoted by white boxes), whole cavities (including the core), and core regions of the north-east and south-east coronal cavities are shown in Figure~\ref{fig:cavity1} (f) and Figure~\ref{fig:cavity2} (f). We find that the non-thermal velocities inside coronal cavities are higher than those in the surrounding streamer regions. In particular, the non-thermal velocities are highest in the core regions of coronal cavities, which are directly above the prominences. However, unlike the intensity and Doppler velocity maps, the non-thermal velocity maps do not show clear concentric semicircular layer structures. The streamer regions near the two coronal cavities have similar values. The non-thermal velocities of north-east cavity and its core region are significantly higher than those of south-east cavity.

\section{Discussion}\label{sec:Discussion}

\subsection{The Doppler Shift Structures inside Coronal Cavities}

SICG observations reveal ring-like Doppler shift patterns inside coronal cavities, which are consistent with previous results. \citet{Schmit2009} first detected Doppler shifts in coherent structures inside coronal cavities using CoMP and EIS observations. They found that the flows showed both red-shift and blue-shift. The speed of the flows ranged from $5$ to $10~\mathrm{km\,s^{-1}}$ with length scales of tens of megameters, and persisted for at least 1 hour. Their results demonstrate that there exist material flows inside coronal cavities. \citet{BS2016} analyzed 66 days of observations of coronal cavities observed by CoMP. They found that most coronal cavities exhibited concentric rings of alternating red-shift and blue-shift, with the magnitude of  Doppler velocities ranging from $4$ to $8~\mathrm{km\,s^{-1}}$, which could persist for several days. The persistence of these flows suggests that they are stable features, not short-term events. Notably, they found that the largest Doppler velocities are typically located just above the prominence. Similarly, the SICG observations reveal the same spatial distribution of the enhanced Doppler shifts.

The consistent Doppler shift patterns observed in different datasets may be attributed to the magnetic field configuration within the coronal cavity. Previous studies modeled coronal cavities as flux ropes \citep{Xia2014, Fan2018, Jenkins2021}. In a low-$\beta$ environment such as the corona, coronal plasma is constrained by magnetic field lines. Therefore, the ring-like Doppler shift structures can be interpreted as plasma flows transported along magnetic field lines inside coronal cavities \citep{Schmit2009, Chen2018,BS2013}. The Doppler structures are not perfectly symmetric, and the largest Doppler velocities are typically located just above the prominence. The results may reflect a combination of complex magnetic field configurations and line-of-sight projection effects. In the simulations by \citet{Xia2014}, the cavity contains two types of magnetic field lines: arched field lines and helical field lines. Arched field lines lie above the flux rope axis. These field lines do not support dense prominence materials and remain largely empty. Concave upward helical field lines lie below the axis. These lower field lines often carry cool and dense prominence plasma. These differences in magnetic field line topology can result in uneven distributions of flow speed. The largest Doppler velocities observed just above the prominence may be related to the magnetic field configuration in the cavity core, where field lines are less twisted and more aligned with the line of sight, enhancing the projected flow velocity. In principle, coronal seismology applied to coronagraph spectroscopic observations can be used to determine both the strength \citep{2020Sci...369..694Y,2024Sci...386...76Y} and the orientation \citep{2020ScChE..63.2357Y,2024Sci...386...76Y} of the coronal magnetic field. In future work, this approach can be applied to diagnose the magnetic structure of coronal cavities.

\subsection{The Enhanced Non-Thermal Velocities and Instabilities}

SICG spectroscopic observations show that line width inside the coronal cavities is significantly larger than that in the surrounding streamer regions. To quantify the non-thermal component, we assume that the ion temperature equals the electron temperature and use the peak formation temperature of \ion{Fe}{14} (about 2~MK) \citep{Habbal2014} to derive the non-thermal velocities. The non-thermal broadening is generally considered to result from the combined effects of differential bulk motion and wave/turbulence motion \citep{Xia2003}. In SICG observations, alternating red- and blue-shifted regions are clearly visible in the Doppler velocity maps, suggesting the presence of complex velocity fields  inside coronal cavities. Therefore, the physical origin of enhanced non-thermal broadening cannot be uniquely determined. To fully disentangle the relative contributions of turbulence and unresolved flows to the observed line broadening, further investigations are required.

If the non-thermal broadening is mainly attributed to wave and/or turbulence motions, SICG observations provide direct quantitative evidence of enhanced plasma fluctuations and turbulence inside coronal cavities, particularly in their core regions. White-light eclipse observations reveal complex internal structures in coronal cavities, including twisted helical structures, small bubbles, and turbulence-like features that appear to fill the volume just above the prominences \citep{Habbal2010c, Druckmuller2014, Habbal2014, Habbal2021}. In addition, these structures at the tip of the prominence also exhibit temporal evolution \citep{Habbal2014}. Coronal cavities are regions where the cold, dense prominence materials interact with the hot, low-density coronal plasma. These regions provide an ideal environment for the development of Kelvin-Helmholtz instability (KHI) and Rayleigh-Taylor instability (RTI) \citep{Ryutova2010, Soler2012, Habbal2014,Druckmuller2014}. Generally speaking, the morphological complexity and dynamical evolution revealed by white-light eclipse observations indicate that complex plasma motions and instabilities may occur within coronal cavities \citep{Druckmuller2014, Habbal2014}. In addition, the SICG spectroscopic observations suggest that MHD waves and/or plasma turbulence exist within coronal cavities, and the waves and turbulence inside coronal cavities are likely stronger than those in the surrounding streamer regions.

\section{Summary}\label{sec:Summary}

SICG provides spectral diagnostics of coronal cavities, offering new insights into the thermodynamic and magnetic properties of these pre-eruption structures. In the present study, we analyzed the characteristics of coronal cavities using new SICG observations, along with images taken by the SDO/AIA. The Doppler shift and non-thermal velocity inside coronal cavities and surrounding streamer structures were analyzed and compared. We find that:

\begin{enumerate}
    \item The coronal cavities exhibit asymmetric, ring-like Doppler shift structures. The maximum Doppler velocities are found inside the core regions and are located just above the prominences. The velocities range from $-12$ to $12~\mathrm{km~s^{-1}}$.
    \item The non-thermal velocities inside coronal cavities are significantly higher than those in the surrounding streamers. In addition, the core regions just above the prominences exhibit the highest non-thermal velocities. In the north-east coronal cavity, the non-thermal velocities of streamer, whole cavity, and core region are $13.87 \pm 2.14~\mathrm{km~s^{-1}}$, $19.12 \pm 3.66~\mathrm{km~s^{-1}}$, and $24.69 \pm 1.86~\mathrm{km~s^{-1}}$, respectively. In the south-east coronal cavity, the non-thermal velocities of streamer, whole cavity, and core region are $14.22 \pm 1.41~\mathrm{km~s^{-1}}$, $16.69 \pm 2.06~\mathrm{km~s^{-1}}$, and $18.98 \pm 2.07~\mathrm{km~s^{-1}}$, respectively.
\end{enumerate}

Our results suggest that MHD waves and/or plasma turbulence exist within coronal cavities, and furthermore, the waves and turbulence inside coronal cavities are likely stronger than those in the surrounding streamer regions. We suggest that the interaction and exchange between the cold, dense prominence materials and the hot, low-density coronal materials are the main drivers of the waves and turbulence inside coronal cavities. These waves and turbulence may contribute to localized plasma heating within the coronal cavities. The study sheds light on the intricate dynamics in prominence-cavity systems. Future work will focus on the temporal evolution and statistical manners of spectroscopic properties within coronal cavities.

\begin{acknowledgments}

The authors gratefully acknowledge the constructive comments and insightful suggestions provided by anonymous referee, which have greatly improved the quality of this work. We acknowledge the use of data from the Chinese Meridian Project. The CHASE mission is supported by China National Space Administration. SDO is a mission of NASA's Living With a Star Program. We gratefully acknowledge discussions with Prof.~Hui Tian. This research is supported by the National Natural Science Foundation of China (42230203 and 12473058).

\end{acknowledgments}

\begin{appendix}\label{appendix}
\counterwithin{figure}{section}
\counterwithin{table}{section}

\section{The Details of Morphology and Temperature of Coronal Cavities}\label{appendix}

\begin{figure*}[ht!]
\plotone{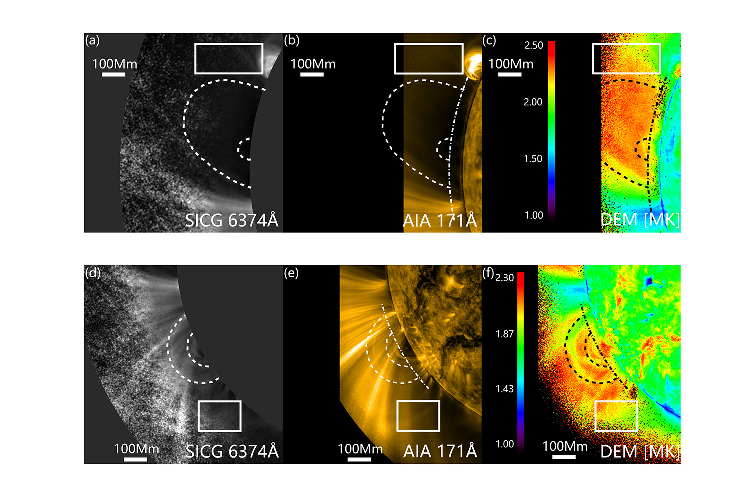}
\caption{The structures on SICG \ion{Fe}{10} 6374~\AA, AIA 171 \AA \ and temperatures of north-east (top row) and south-east (bottom row) coronal cavities. The intensity maps of the SICG \ion{Fe}{10} 6374~\AA\ central wavelength are shown in panels (a) and (d). The AIA 171 \AA \ images are shown in panels (b) and (e). The temperature maps derived from DEM method are shown in panels (c) and (f). In panels (b), (c), (e) and (f), the positions of the SICG occulter are indicated by white/black dash-dotted lines. The boundaries of the coronal cavities and core regions are marked with dashed lines, and the white boxes denote streamer regions near the two cavities.}
\label{morphology and temperature}
\end{figure*}

The intensity maps of SICG \ion{Fe}{10} 6374~\AA, AIA 171~\AA\, and temperatures derived from DEM analysis are shown in Figure A.1.  The coronal cavities are distinctly visible in SICG \ion{Fe}{14} 5303~\AA\ (Figure~\ref{fig:cavity1} (a) and Figure~\ref{fig:cavity2} (a)) and AIA 193~\AA\ (Figure~\ref{fig:cavity1} (d) and Figure~\ref{fig:cavity2} (d)), both of which correspond to emission lines formed at coronal temperatures around 2~MK. In contrast, the north-east cavity is barely visible in the \ion{Fe}{10} 6374~\AA\ image (Figure~\ref{morphology and temperature} (a)) and the AIA 171~\AA\ image (Figure~\ref{morphology and temperature} (b)). The south-east cavity is obscured by the stronger overlying fan-shaped structures (Figure~\ref{morphology and temperature} (d) and (e)). The combined SICG and AIA observations reflect plasma temperature inside coronal cavities, indicating that the temperature inside coronal cavities is nearly 2~MK. The temperatures derived by DEM method (see Figure~\ref{morphology and temperature} (c) and (f)) support the indication. The average temperatures of coronal cavities are about $2.2 \pm 0.2$~MK and $2.0 \pm 0.1$~MK for north-east and south-east cavities.

The temperatures of coronal cavities measured in the present work are consistent with previous studies, which generally indicate that cavities are relatively hot structures with characteristic temperatures of 1.5-2~MK. Using Yohkoh/SXT observations, \citet{Hudson1999} found cavity temperatures in the range of 1.63-1.84~MK. Hinode/XRT observations analyzed by \citet{Reeves2012} gave central cavity temperatures of 1.75, 1.7, and 2.0~MK on three different days, suggesting progressive heating over time. Using Hinode/EIS spectroscopic data, \citet{Kucera2012} diagnosed the temperatures of the coronal cavity and the surrounding streamer region. They found no significant temperature difference between the cavity and the streamer region, with both having temperatures in the range of 1.4-1.7 MK. A DEM analysis by \citet{2009SoPh..256...73V} using EUVI data from STEREO A and B showed that the EM distribution in coronal cavity is broader than that in the surrounding streamer region. The cavity contain more emission measure above 2~MK than the streamer plasma. A statistical DEM study of 33 coronal cavities using AIA data obtained temperatures between 1.67 and 2.15~MK \citep{BS2019}. Image observations taken in Fe~X 637.4~nm, Fe~XI 789.2~nm, Fe~XIII 1074.7~nm, and Fe~XIV 530.3~nm during a solar eclipse indicate that the coronal cavity emission is mainly dominated by plasma at 2~MK or even higher temperatures \citep{Habbal2010c}.

The north-east coronal cavity appears larger in size and extends to a greater height compared to the south-east cavity.  In addition, the prominence below the north-east cavity is also located higher than that beneath the south-east cavity. This observation is consistent with the statistical findings of \citet{BS2016}, who reported that higher prominences are generally associated with higher-lying coronal cavities. This correlation is likely rooted in the magnetic field configuration of the prominence--cavity system. A higher prominence typically implies a larger-scale magnetic flux rope or sheared arcade, which can support the surrounding hot, low-density cavity material at greater altitudes. Moreover, the difference in cavity height and size may reflect distinct evolutionary stages \citep{Gibson2006}.

\end{appendix}


\bibliography{sample701}{}
\bibliographystyle{aasjournalv7}



\end{sloppypar}
\end{document}